\documentclass[final,1p,times,twocolumn]{elsarticle}
\usepackage{lineno,hyperref,amsmath,amsthm,amssymb,amsfonts,ragged2e,color,subfig}
\journal{``Journal of Applied Mechanics and Technical Physics"}
\begin{document}
\begin{frontmatter}
\title{Envelope solitons in double pair plasmas}
\author{N. K. Tamanna$^{*,1}$, N. A. Chowdhury$^{**,1}$, A. Mannan$^{1,2}$, and A. A. Mamun$^{1}$}
\address{$^{1}$Department of Physics, Jahangirnagar University, Savar, Dhaka-1342, Bangladesh\\
$^{2}$Institut f\"{u}r Mathematik, Martin Luther Universit\"{a}t Halle-Wittenberg, Halle, Germany\\
Email: $^*$tamanna1995phy@gmail.com, $^{**}$nurealam1743phy3@gmail.com}
\begin{abstract}
A double pair plasma system containing cold inertial positive and negative ions, and
inertialess super-thermal electrons and positrons is considered. The standard nonlinear
Schr\"{o}dinger equation is derived by using the reductive perturbation method to
investigate the nonlinear dynamics of the ion-acoustic waves (IAWs) as well as their
modulation instability. It is observed that the ion-acoustic dark (bright) envelope
solitons are formed for modulationally stable (unstable) plasma region, and that the
presence of highly dense super-thermal electrons and positrons enhances (reduces)
this unstable (stable) region. It is also found that the effect of super-thermality
of electron or positron species causes to increase the nonlinearity, and to fasten
the formation of the bright envelope solitons. These results are applicable to both space and
laboratory plasma systems for understanding the propagation of localized electrostatic disturbances.
\end{abstract}
\end{frontmatter}
\section{Introduction}
\label{1sec:Introduction}
Double pair plasmas are considered as fully ionized gases consisting of positively and negatively charged ions
as well as an electron and positron having equal mass and opposite charge, and have been
observed to exist in space environments, viz., upper regions of Titan's atmosphere \cite{El-Labany2012a},
chromosphere, solar wind, D-region ($\rm H^+, O_2^-$) and F-region ($\rm H^+, H^-$) of the
Earth's ionosphere \cite{Elwakil2010}, cometary comae \cite{Chaizy1991,Chowdhury2017},
and laboratory situations, viz., neutral beam sources \cite{Bacal1979}, plasma processing
reactors \cite{Gottscho1986}, and Fullerene ($\rm C^+, C^-$) \cite{Sabry2008}, etc.

The super-thermal features of fast particles in a plasma system are measured by
the deviation of the plasma particles distribution from the well known Maxwellian
distribution, and this deviation is expressed by the super-thermal parameter ($\kappa$) in
a super-thermal $\kappa$-distribution \cite{Vasyliunas1968,Sultana2011,Ghosh2012,Chatterjee2011,Saha2014}.
The deviation between Kappa and Maxwellian distribution is negligible for large values
of $\kappa$ ($\kappa\rightarrow\infty$). The super-thermality of the plasma species
leads an effective change on the electron-acoustic waves (EAWs) \cite{Sultana2011}, positron-acoustic waves (PAWs), ion-acoustic (IA) waves (IAWs),
IA solitary waves (IASWs) \cite{Ghosh2012}, and  IA double layers (IA-DLs) \cite{El-Labany2012b}.
Sultana and Kourakis studied EAWs in super-thermal plasmas to examine the stability criterion of
the EAWs. Ghosh \textit{et al.} \cite{Ghosh2012} considered unmagnetized three components
plasma having super-thermal electrons and positrons, and observed the velocity of the
solitary waves decreases with increasing $\kappa$. Chatterjee and Ghosh \cite{Chatterjee2011} investigated head-on-collision
of IASWs in a three components plasma medium. Saha \textit{et al.} \cite{Saha2014} reported dynamic
properties of the IASWs by considering super-thermal electrons and positrons, and found that the amplitude of the IASWs
decreases while width increases with increasing $\kappa$.

The occurrence of the electrostatic bright and dark envelope solitons,
and modulational instability (MI) of the carrier waves in any plasma medium are demonstrated
by the standard nonlinear Schr\"{o}dinger equation (NLSE) \cite{C1,C3,C6,Gharaee2011,Chowdhury2018,Alinejad2014,Eslami2011,Ahmed2018,Kourakis2003,Kourakis2005,C2,C4,C5}.
Gharaee \textit{et al.} \cite{Gharaee2011} have reported the MI of the IAWs in an electron-ion plasma system
and obtained that the critical frequency of the MI of IAWs reduces with the presence of the super-thermal electrons.
Chowdhury \textit{et al.} \cite{Chowdhury2018} have demonstrated the formation of IA envelope solitons in
a quantum plasma, and have found that the magnitude of the amplitude of the IA envelope solitons remains
constant against the variation of different plasma parameters. Alinejad \textit{et al.} \cite{Alinejad2014}
have investigated the MI of IAWs in a three components plasma medium with $\kappa$-distributed electrons, and
have observed that the $k_c$ decreases with $\kappa$. Eslami \textit{et al.} \cite{Eslami2011} have
investigated the stability criterion of IAWs in an electron-positron-ion (e-p-i) plasma and observed that the number density and temperature of the
positron significantly modify the nature of the stability conditions of IAWs in an e-p-i plasma.
Ahmed \textit{et al.} \cite{Ahmed2018} have studied MI of IAWs in an e-p-i plasma
medium in presence of $\kappa$-distributed electrons and positrons, and have found that
the critical waves number ($k_c$) decreases with an increase in the value of $\kappa$.
El-Labany \textit{et al.} \cite{El-Labany2012a} have examined the stability of the IAWs in a three components pair ion plasma medium.

Recently, Dubinov \textit{et al.} \cite{Dubinov2006} have investigated IASWs in a symmetric pair-ion plasma (PIP).
El-Labany \textit{et al.} \cite{El-Labany2012b} observed IA-DLs in a three components double ion plasma
medium as well as how the positive and negative ions organize the critical point.
Dubinov \textit{et al.} \cite{Dubinov2011} examined electrostatic baryonic waves in an ambiplasma medium containing
protons, antiprotons, electrons and positrons.
Jannat \textit{et al.} \cite{Jannat2016} studied Gardner Solitons by considering
four components PIP model having inertial positive and negative ions as well as inertialess electrons and
positrons. In our present paper, we have extended the work of Jannat \textit{et al.} \cite{Jannat2016}
by considering a mathematical model regarding four components PIP model consisting of
positive and negative ion as well as electron and positron featuring super-thermal $\kappa$-distribution, and
will study the MI of IAWs and the configuration of the electrostatic IA bright and dark envelope solitons.

The outline of the paper is as follows: The governing equations describing our plasma
model are presented in Section \ref{1sec:Governing Equations}. A NLSE is derived in
Section \ref{1sec:Derivation of the NLSE}. Modulational instability and envelope solitons
are given in Section \ref{1sec:Modulational instability and envelope solitons}. A brief conclusion is,
finally, provided in Section \ref{1sec:Conclusion}.
\newpage
\section{Governing Equations}
\label{1sec:Governing Equations}
We consider a four components plasma system consisting of inertial negative ions, positive ions and inertialess electrons and positrons.
At equilibrium, the quasi-neutrality condition can be expressed as $Z_+n_{+0}+n_{p0}=Z_-n_{-0}+n_{e0}$; where
$n_{+0}$, $n_{p0}$, $n_{-0}$, and $n_{e0}$ are the equilibrium number densities of positive ions, super-thermal positrons,
negative ions, and super-thermal electrons, respectively.
Now, the basic set of normalized equations can be written in the form
\begin{eqnarray}
&&\hspace*{-4cm}\frac{\partial n_-}{\partial t}+\frac{\partial}{\partial x}(n_-u_-)=0,
\label{1eq:1}\\
&&\hspace*{-4cm}\frac{\partial u_-}{\partial t}+u_-\frac{\partial u_-}{\partial x}=\gamma_1\frac{\partial\phi}{\partial x},
\label{1eq:2}\\
&&\hspace*{-4cm}\frac{\partial n_+}{\partial t}+\frac{\partial}{\partial x}(n_+u_+)=0,
\label{1eq:3}\\
&&\hspace*{-4cm}\frac{\partial u_+}{\partial t}+u_+\frac{\partial u_+}{\partial x}=-\frac{\partial\phi}{\partial x},
\label{1eq:4}\\
&&\hspace*{-4cm}\frac{\partial^2\phi}{\partial x^2}=\gamma_2n_e-\gamma_3n_p+(1-\gamma_2+\gamma_3)n_--n_+,
\label{1eq:5}
\end{eqnarray}
where $n_-$ and $n_+$ are the negative and positive ion number density normalized
by their equilibrium value $n_{-0}$ and $n_{+0}$, respectively; $u_-$ and $u_+$ are
the negative and positive ion fluid speed normalized by wave speed $C_+=(Z_+k_BT_e/m_+)^{1/2}$
(with $T_e$ being the temperature super-thermal electron, $m_+$ being the positive ion mass,
and $k_B$ being the Boltzmann constant); $\phi$ is the electrostatic wave potential
normalized by $k_BT_e/e$ (with $e$ being the magnitude of single electron charge); the
time and space variables are normalized by $\omega_{p+}^{-1}=(m_+/4\pi e^2 Z_+^2n_{+0})^{1/2}$
and $\lambda_{D+}=(k_BT_e/4\pi e^2 Z_+n_{+0})^{1/2}$, respectively; and $\gamma_1=Z_-m_+/Z_+m_-$,
$\gamma_2=n_{e0}/Z_+n_{+0}$, and $\gamma_3=n_{p0}/Z_+n_{+0}$.
Now, the expressions for electron and positron number
density obeying $\kappa$-distribution are, respectively, given by \cite{Vasyliunas1968,Sultana2011,Chowdhury2017}
\begin{eqnarray}
&&\hspace*{-6.5cm}n_e=\left[1 - \frac{\phi}{\kappa-3/2}\right]^{-\kappa+1/2},
\label{1eq:6}\\
&&\hspace*{-6.5cm}n_p=\left[1 +\frac{\gamma_4\phi}{\kappa-3/2}\right]^{-\kappa+1/2},
\label{1eq:7}
\end{eqnarray}
where $\gamma_4 =T_e/T_p$. Here, the parameter $\kappa$ represents the super-thermal property of the electrons and positrons.
Now, by substituting \eqref{1eq:6} and \eqref{1eq:7} into \eqref{1eq:5}, and expanding up to
third order of $\phi$, we get
\begin{eqnarray}
&&\hspace*{-1cm}\frac{\partial^2\phi}{\partial x^2} + n_{+}+\gamma_3= \gamma_2 +(1-\gamma_2+\gamma_3)n_{-}+ M_1\phi + M_2\phi^2 + M_3\phi^3 + \cdot\cdot\cdot,
\label{1eq:8}
\end{eqnarray}
where
\begin{eqnarray}
&&\hspace*{-1.3cm}M_1=\frac{(\gamma_2+\gamma_3\gamma_4)(2\kappa-1)}{(2\kappa-3)},~~~~~M_2=\frac{(\gamma_2-\gamma_3\gamma_4^2)(2\kappa-1)(2\kappa+1)}{2(2\kappa-3)^2},
\nonumber\\
&&\hspace*{-1.3cm}M_3=\frac{(\gamma_2+\gamma_3\gamma_4^3)(2\kappa-1)(2\kappa+1)(2\kappa+3)}{6(2\kappa-3)^3}.
\nonumber\
\end{eqnarray}
We note that the term on the right hand side of the Eq. \eqref{1eq:8} is the contribution of
super-thermal electrons and positrons.
\section{Derivation of the NLSE}
\label{1sec:Derivation of the NLSE}
\begin{figure}[t!]
\centering
\includegraphics[width=80mm]{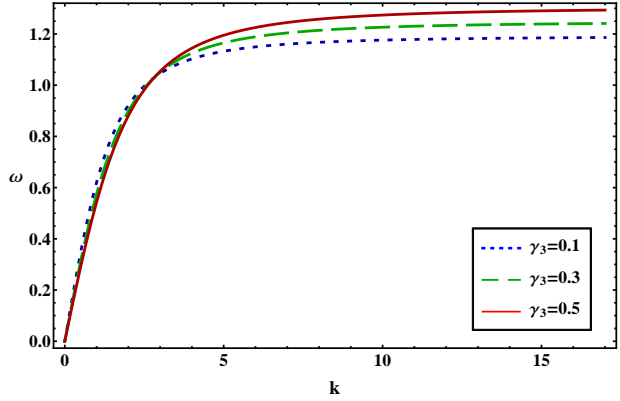}
\caption{Plot of $\omega$ vs $k$ for different values of $\gamma_3$ when $\gamma_1=0.7$, $\gamma_2=0.5$, $\gamma_4=1.2$, and $\kappa=1.8$.}
\label{1Fig:F1}
\end{figure}

To study the MI of the IAWs, we want to derive the NLSE by employing the reductive perturbation method (RPM)
and for that case, we can write the stretched co-ordinates in the form \cite{Kourakis2003,Kourakis2005}
\begin{eqnarray}
&&\hspace*{-8.8cm}\xi={\epsilon}(x-v_g t),
\label{1eq:9}\\
&&\hspace*{-8.8cm}\tau={\epsilon}^2 t,
\label{1eq:10}
\end{eqnarray}
where $v_g$ is the group velocity and $\epsilon$ is a small parameter. Then, we can write the dependent variables as \cite{Kourakis2003,Kourakis2005}
\begin{eqnarray}
&&\hspace*{-4.6cm}n_-=1+\sum_{m=1}^{\infty}\epsilon^{(m)}\sum_{l=-\infty}^{\infty} n_{-l}^{(m)}(\xi,\tau)~\mbox{exp}[i l(kx-\omega t)],
\label{1eq:11}\\
&&\hspace*{-4.6cm}u_-=\sum_{m=1}^{\infty}\epsilon^{(m)}\sum_{l=-\infty}^{\infty} u_{-l}^{(m)}(\xi,\tau)~\mbox{exp}[i l(kx-\omega t)],
\label{1eq:12}\\
&&\hspace*{-4.6cm}n_+=1+\sum_{m=1}^{\infty}\epsilon^{(m)}\sum_{l=-\infty}^{\infty} n_{+l}^{(m)}(\xi,\tau)~\mbox{exp}[i l(kx-\omega t)],
\label{1eq:13}\\
&&\hspace*{-4.6cm}u_+=\sum_{m=1}^{\infty}\epsilon^{(m)}\sum_{l=-\infty}^{\infty} u_{+l}^{(m)}(\xi,\tau)~\mbox{exp}[i l(kx-\omega t)],
\label{1eq:14}\\
&&\hspace*{-4.6cm}\phi=\sum_{m=1}^{\infty}\epsilon^{(m)}\sum_{l=-\infty}^{\infty} \phi_{l}^{(m)}(\xi,\tau)~\mbox{exp}[i l(kx-\omega t)].
\label{1eq:15}\
\end{eqnarray}
The derivative operators can be written as
\begin{eqnarray}
&&\hspace*{-8cm}\frac{\partial}{\partial t}\rightarrow\frac{\partial}{\partial t}-\epsilon v_g \frac{\partial}{\partial\xi}
+\epsilon^2\frac{\partial}{\partial\tau},
\label{1eq:16}\\
&&\hspace*{-8cm}\frac{\partial}{\partial x}\rightarrow\frac{\partial}{\partial x}+\epsilon\frac{\partial}{\partial\xi}.
\label{1eq:17}
\end{eqnarray}
Now, by substituting \eqref{1eq:9}-\eqref{1eq:17} into \eqref{1eq:1}-\eqref{1eq:4}, and \eqref{1eq:8}, and
collecting the terms containing $\epsilon$, the first order ($m=1$ with $l=1$) reduced equations can be written as
\begin{eqnarray}
&&\hspace*{-9cm}n_{-1}^{(1)}=-\frac{\gamma_1 k^2}{\omega^2}\phi_1^{(1)},
\label{1eq:18}\\
&&\hspace*{-9cm}u_{-1}^{(1)}=-\frac{\gamma_1 k}{\omega}\phi_1^{(1)},
\label{1eq:19}\\
&&\hspace*{-9cm}n_{+1}^{(1)}=\frac{k^2}{\omega^2}\phi_1^{(1)},
\label{1eq:20}\\
&&\hspace*{-9cm}u_{+1}^{(1)}=\frac{k}{\omega}\phi_1^{(1)},
\label{1eq:21}\
\end{eqnarray}
these relation provides the dispersion relation for IAWs
\begin{eqnarray}
&&\hspace*{-7cm}\omega^2=\frac{k^2[1+\gamma_1(1-\gamma_2+\gamma_3)]}{k^2+M_1}.
\label{1eq:22}
\end{eqnarray}
We have numerically analysed Eq. \eqref{1eq:22} to examine the linear dispersion properties of IAWs for different values of the
$\gamma_3$. The results are displayed in Fig. \ref{1Fig:F1}, which shows that (i) in the short wavelength limit,
the dispersion curves become saturated and the maximum frequency of the IAWs is equal to the positive ion plasma frequency
($\omega_{p+}$); (ii) in the long wavelength limit, the angular frequency of the IAWs linearly increases with the wave number k;
(iii) the nature of the IAWs are, therefore, similar to other kinds of acoustic-type waves (i.e. EAWs and
PAWs) but with different time and length scales. The angular frequency of IAWs, sometimes, can be found
in the form of biquadratic equation by considering the adiabatic pressure \cite{Dubinov2011,Saberian2017,Kourakis2006,Esfandyri-Kalejahi2006a,Esfandyri-Kalejahi2006b} in the
governing equations specially in the momentum equations. As biquadratic equation has two
possible positive solutions corresponding to the positive and negative sign. Then, it may be considered that the positive
(negative) sign represents first (slow) IA modes \cite{Saberian2017}. The fast IA mode also corresponds to the case in which both ion species oscillate in phase with electrons and positrons. On the other hand,
the slow IA mode corresponds to the case in which only one ion oscillate in phase with electrons and positrons
while other ion species are in anti-phase with them \cite{Saberian2017}. If anyone considers the dynamics of the pair-ion
without considering adiabatic pressure \cite{El-Labany2012a,Elwakil2010,Abdelsalam2011,El-Labany2011} in the governing equations then it may not be possible to find the angular frequency in the form of biquadratic equation.
The second-order ($m=2$ with $l=1$) equations are given by
\begin{eqnarray}
&&\hspace*{-6.0cm}n_{-1}^{(2)}=-\frac{\gamma_1k^2}{\omega^2}\phi_1^{(2)}-\frac{2i\gamma_1k(v_gk-\omega)}{\omega^3}\frac{\partial \phi_1^{(1)}}{\partial\xi},
\label{1eq:23}\\
&&\hspace*{-6.0cm}u_{-1}^{(2)}=-\frac{\gamma_1k}{\omega}\phi_1^{(2)}-\frac{i\gamma_1(v_gk-\omega)}{\omega^2}\frac{\partial \phi_1^{(1)}}{\partial\xi},
\label{1eq:24}\\
&&\hspace*{-6.0cm}n_{+1}^{(2)}=\frac{k^2}{\omega^2}\phi_1^{(2)}+\frac{2ik(v_gk-\omega)}{\omega^3}\frac{\partial \phi_1^{(1)}}{\partial\xi},
\label{1eq:25}\\
&&\hspace*{-6.0cm}u_{+1}^{(2)}=\frac{k}{\omega}\phi_1^{(2)}+\frac{i(v_gk-\omega)}{\omega^2}\frac{\partial \phi_1^{(1)}}{\partial\xi},
\label{1eq:26}\
\end{eqnarray}
with the compatibility condition
\begin{eqnarray}
&&\hspace*{-6cm}v_g=\frac{\partial \omega}{\partial k}=\frac{\omega[1+\gamma_1(1-\gamma_2+\gamma_3)-\omega^2]}{k[1+\gamma_1(1-\gamma_2+\gamma_3)                             ]}.
\label{1eq:27}\
\end{eqnarray}
The coefficients of $\epsilon$ for $m=2$ with $l=2$ provide the second
order harmonic amplitudes which are found to be proportional to $|\phi_1^{(1)}|^2$
\begin{eqnarray}
&&\hspace*{-9.4cm}n_{_-2}^{(2)}=M_4|\phi_1^{(1)}|^2,
\label{1eq:28}\\
&&\hspace*{-9.4cm}u_{-2}^{(2)}=M_5 |\phi_1^{(1)}|^2,
\label{1eq:29}\\
&&\hspace*{-9.4cm}n_{+2}^{(2)}=M_6|\phi_1^{(1)}|^2,
\label{1eq:30}\\
&&\hspace*{-9.4cm}u_{+2}^{(2)}=M_7 |\phi_1^{(1)}|^2,
\label{1eq:31}\\
&&\hspace*{-9.4cm}\phi_{2}^{(2)}=M_8 |\phi_1^{(1)}|^2,
\label{1eq:32}\
\end{eqnarray}
where
\begin{eqnarray}
&&\hspace*{-3cm}M_4=\frac{3\gamma_1^2k^4-2\gamma_1\omega^2k^2M_8}{2\omega^4},~~~~~M_5=\frac{\gamma_1^2k^3-2\gamma_1\omega^2k M_8}{2\omega^3},
\nonumber\\
&&\hspace*{-3cm}M_6=\frac{3k^4+2\omega^2k^2M_8}{2\omega^4},~~~~~~M_7=\frac{k^3+2\omega^2kM_8}{2\omega^3},
\nonumber\\
&&\hspace*{-3cm}M_8=\frac{2M_2\omega^4-3k^4+3\gamma_1^2k^4(1-\gamma_2+\gamma_3)}{2\omega^2k^2-2\omega^4(4k^2+M_1)+2\gamma_1\omega^2k^2(1-\gamma_2+\gamma_3)}.
\nonumber\
\end{eqnarray}
Now, we consider the expression for ($m=3$ with $l=0$) and ($m=2$ with $l=0$),
which leads the zeroth harmonic modes. Thus, we obtain
\begin{eqnarray}
&&\hspace*{-9cm}n_{-0}^{(2)}=M_{9}|\phi_1^{(1)}|^2,
\label{1eq:33}\\
&&\hspace*{-9cm}u_{-0}^{(2)}=M_{10}|\phi_1^{(1)}|^2,
\label{1eq:34}\\
&&\hspace*{-9cm}n_{+0}^{(2)}=M_{11}|\phi_1^{(1)}|^2,
\label{1eq:35}\\
&&\hspace*{-9cm}u_{+0}^{(2)}=M_{12}|\phi_1^{(1)}|^2,
\label{1eq:36}\\
&&\hspace*{-9cm}\phi_0^{(2)}=M_{13}|\phi_1^{(1)}|^2,
\label{1eq:37}\
\end{eqnarray}
where
\begin{eqnarray}
&&\hspace*{-0cm}M_{9}=\frac{\omega\gamma_1^2k^2+2\gamma_1^2k^3v_g-\gamma_1\omega^3M_{13}}{v_g^2\omega^3},~~~~~~~M_{10}=\frac{\gamma_1^2k^2-\gamma_1\omega^2 M_{13}}{v_g\omega^2},
\nonumber\\
&&\hspace*{-0cm}M_{11}=\frac{\omega k^2+2v_gk^3+\omega^3 M_{13}}{v_g^2\omega^3},~~~~~~~M_{12}=\frac{k^2+\omega^2 M_{13}}{v_g\omega^2},
\nonumber\\
&&\hspace*{-0cm}M_{13}=\frac{2M_2v_g^2\omega^3+\omega\gamma_1^2k^2(1-\gamma_2+\gamma_3)+2\gamma_1^2k^3v_g(1-\gamma_2+\gamma_3)-k^2\omega-2k^3v_g}
{\omega^3[1+\gamma_1(1-\gamma_2+\gamma_3)-M_1v_g^2]}.
\nonumber\
\end{eqnarray}
Finally, the third harmonic modes ($m=3$) and ($l=1$) with the help of \eqref{1eq:18}$-$\eqref{1eq:37},
give a set of equations, which can be reduced to the following NLSE:
\begin{eqnarray}
&&\hspace*{-6.7cm}i\frac{\partial \Phi}{\partial\tau}+P\frac{\partial^2\Phi}{\partial\xi^2}+Q\mid\Phi\mid^2\Phi=0,
\label{1eq:38}
\end{eqnarray}
where $\Phi=\phi_1^{(1)}$ for simplicity. The dispersion ($P$) and nonlinear ($Q$) coefficients have the following form, respectively:
\begin{eqnarray*}
&&\hspace*{-0.5cm}P=\frac{3v_g}{2}\Big[\frac{v_g}{\omega}-\frac{1}{k}\Big],\\
&&\hspace*{-0.5cm}Q=\frac{\omega^3}{2k^2[1+\gamma_1(1-\gamma_2+\gamma_3)]}\times\Big[3M_3+2M_2(M_8+M_{13})-\frac{k^2(M_6+M_{11})}{\omega^2}\nonumber\\
&&\hspace*{0.0cm}-\frac{2k^3(M_7+M_{12})}{\omega^3}-\frac{\gamma_1k^2(1-\gamma_2+\gamma_3)(M_4+M_9)}{\omega^2}-\frac{2\gamma_1k^3(1-\gamma_2+\gamma_3)(M_5+M_{10})}{\omega^3}\Big].
\end{eqnarray*}
It may be noted here that both $P$ and $Q$ are function of various
plasma parameters such as $\gamma_1$, $\gamma_2$, $\gamma_3$, $\gamma_4$, $\kappa$, and $k$.
So, all the plasma parameters are used to maintain
the nonlinearity and the dispersion properties of the PIP.
\begin{figure}
\centering
\begin{minipage}{.45\linewidth}
\includegraphics[width=\linewidth]{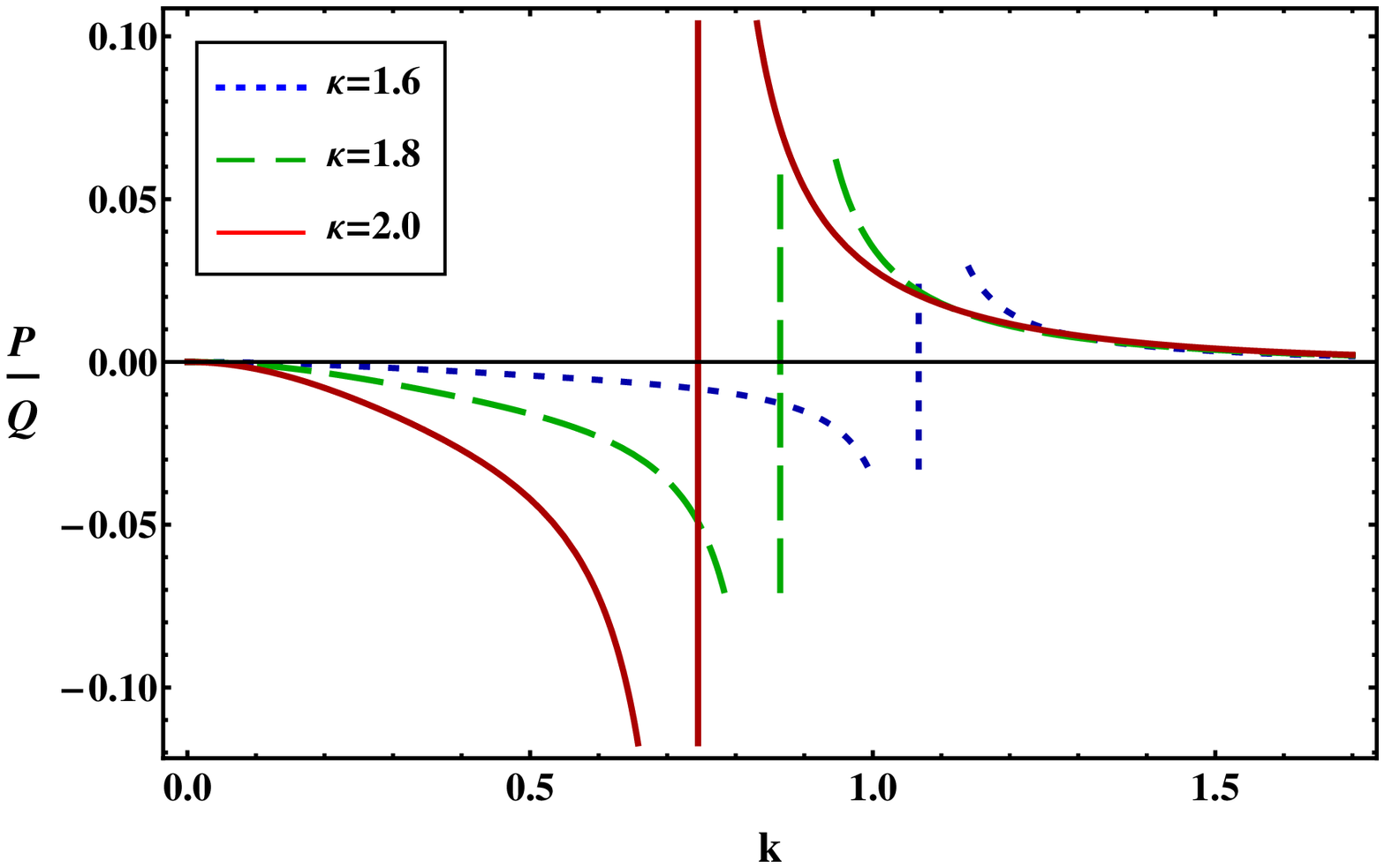}
 \Large{(a)}
\end{minipage}
\hspace{.05\linewidth}
\begin{minipage}{.45\linewidth}
\includegraphics[width=\linewidth]{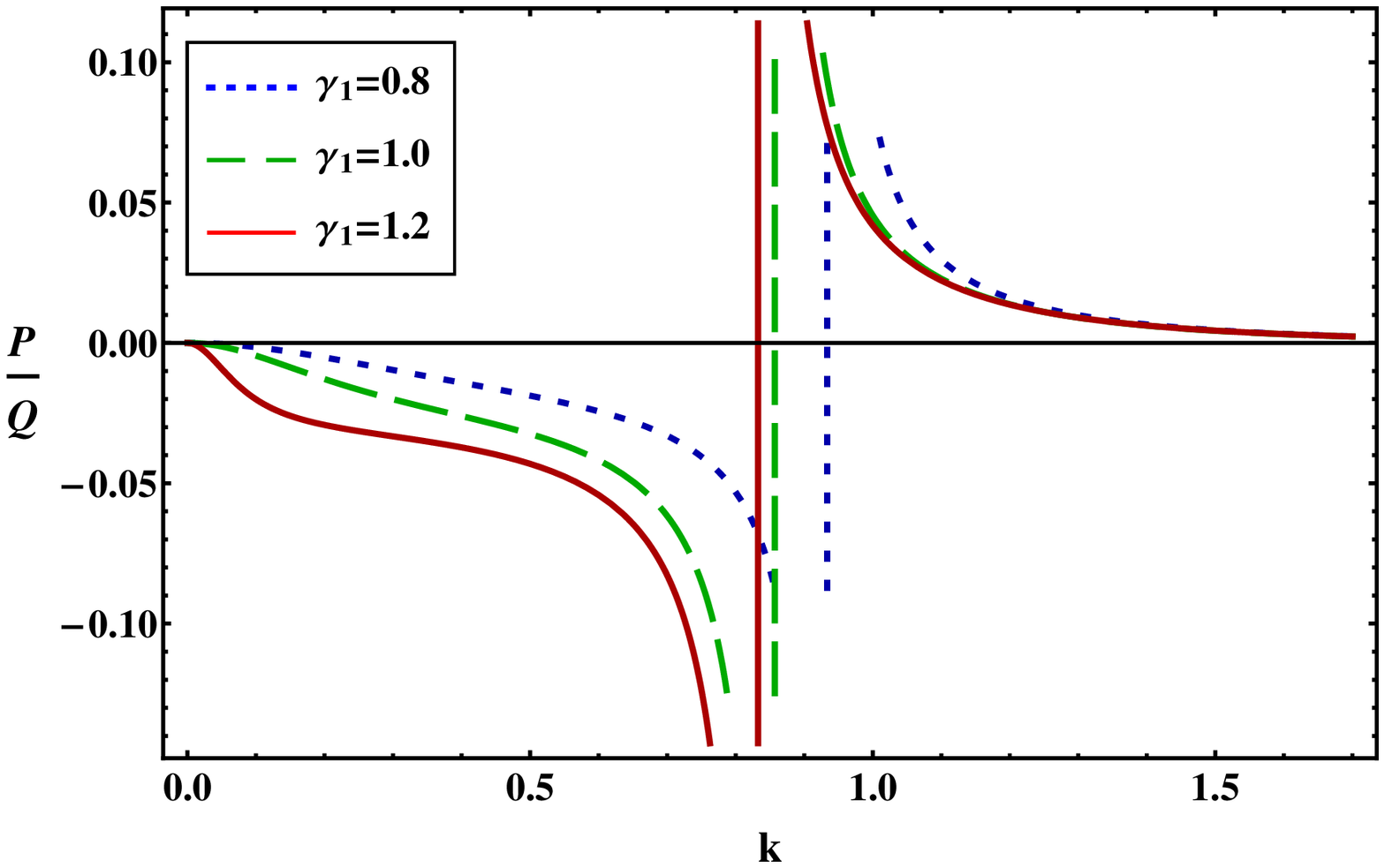}
 \Large{(b)}
\end{minipage}
\caption{Showing the variation of $P/Q$ with $k$ (a) for $\kappa$ and (b) for $\gamma_1$,
when other plasma parameters are $\gamma_2=0.5$, $\gamma_3=0.4$, and $\gamma_4=1.2$.}
\label{1Fig:F2}
\end{figure}
\begin{figure}
\centering
\begin{minipage}{.45\linewidth}
\includegraphics[width=\linewidth]{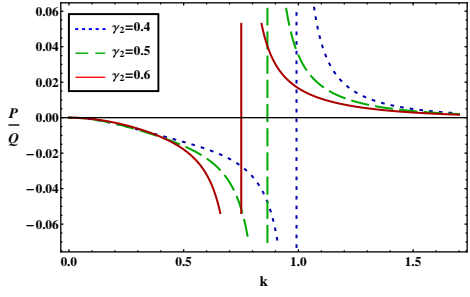}
 \Large{(a)}
\end{minipage}
\hspace{.05\linewidth}
\begin{minipage}{.45\linewidth}
\includegraphics[width=\linewidth]{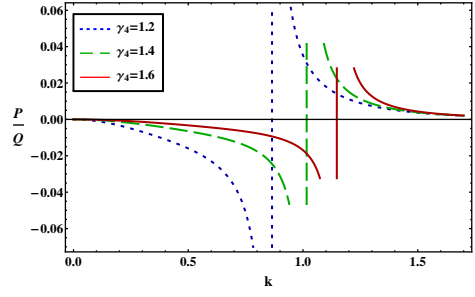}
 \Large{(b)}
\end{minipage}
\caption{Showing the variation of $P/Q$ with $k$ (a) for $\gamma_2$ and (b) for $\gamma_4$,
when other plasma parameters are $\gamma_1=0.7$, $\gamma_3=0.4$, and $\kappa=1.8$.}
\label{1Fig:F3}
\end{figure}
\begin{figure}
\centering
\begin{minipage}{.45\linewidth}
\includegraphics[width=\linewidth]{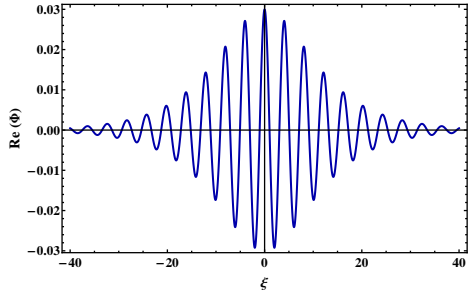}
 \Large{(a)}
\end{minipage}
\hspace{.05\linewidth}
\begin{minipage}{.45\linewidth}
\includegraphics[width=\linewidth]{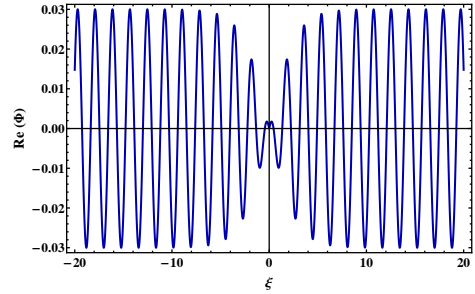}
 \Large{(b)}
\end{minipage}
\caption{(a) Bright envelope solitons for $k=1.0$ and (b) dark envelope solitons for $k=0.2$,
when other plasma parameters are $\tau=0$, $\psi_0=0.0009$, $U=0.3$, $\Omega_0=0.4$, $\kappa=1.8$,
$\gamma_1=0.7$, $\gamma_2=0.5$, $\gamma_3=0.4$, and $\gamma_4=1.2$.}
\label{1Fig:F4}
\end{figure}
\section{Modulational instability and envelope solitons}
\label{1sec:Modulational instability and envelope solitons}
When $P$ and $Q$ are opposite sign then they offer a modulationally stable region for IAWs and allow to
generate electrostatic dark envelope solitions. On the other hand, when both $P$ and $Q$ are of the same sign
then they offer a modulationally unstable region for IAWs and allow to generate electrostatic bright envelope
solitions \cite{Gharaee2011,Chowdhury2018,Alinejad2014,Eslami2011,Ahmed2018,Kourakis2003,Kourakis2005}. The plot of $P/Q$ against $k$ yields
stable and unstable domains for the DAWs. The point, at which transition of $P/Q$ curve intersect with $k$-axis,
is known as threshold or critical wave number $k$ ($=k_c$).

A clear idea about the effects of $\kappa$ on the stable and unstable regions for IAWs can be
observed from Fig. \ref{1Fig:F2}(a), and the outcomes from this figure can be expressed as (i) there
is a stable region (i.e., $k<k_c$) for IAWs in the figure, and this stable region allows to generate the electrostatic dark
envelope solitons; (ii) similarly, there is a unstable region (i.e., $k>k_c$) for IAWs in the figure, and
this unstable region allows to generate the electrostatic bright envelope solitons; (iii) the stable region for the IAWs
decreases with the increase in the value of $\kappa$.
The variation of $P/Q$ with $k$ for different values of $\gamma_1$ can be shown
in Fig. \ref{1Fig:F2}(b) which provides the information about the effects of the
mass of the positive ($m_+$) and negative ($m_-$) ions as well as their charge state of the
positive ($Z_+$) and negative ($Z_-$) ions, and it is clear from this figure
that (i) an increase in the value of $\gamma_1$ can shift the $k_c$ to a smaller
value of $k$ while a decrease in the value of $\gamma_1$ can shift the $k_c$ to
a larger value of $k$; (ii) a massive positive (negative) ion allows to generate
electrostatic bright envelope solitons associated with IAWs for small (large) value of $k$ by
reducing (increasing) $k_c$ (via $\gamma_1=Z_-m_+/Z_+m_-$).

Figure \ref{1Fig:F3}(a) indicates the deviation in the stable and unstable regions of IAWs according to
the variation of $\gamma_2$, and it can be seen from this figure that (i) an increase in the value of  $\gamma_2$
is responsible for a fast destabilization of the IAWs as well as allows to generate electrostatic bright
envelope solitons for small value of $k$; (ii) the presence of excess super-thermal electron would reduce
the $k_c$ to a lower value while the presence of excess inertial positive ion would lead the $k_c$ to a
higher value for a constant value of $Z_+$; (iii) highly dense super-thermal electrons offer bright
envelope solitons (i.e., $k>k_c$) associated IAWs for small $k$ while excess inertial
positive ions offer bright envelope solitons (i.e., $k>k_c$) associated IAWs for
large $k$  (via $\gamma_2=n_{e0}/Z_+n_{+0}$).
Figure \ref{1Fig:F3}(b) reflects the temperature effect of the electron and positron (via $\gamma_4=T_e/T_p$)
in recognizing the stable and unstable regions of IAWs as well as the formation of bright and dark envelope
solitons, and it is obvious from this figure that (i) the stable region for IAWs increases with increasing
value of $\gamma_4$; (ii) the temperature of electron maximizes (minimizes) the $k_c$ as well as modulationally
stable region of IAWs.

The bright (when $P/Q>0$) and dark (when $P/Q<0$) envelope soliton solutions, respectively,
can be written as \cite{Kourakis2003,Kourakis2005,Sultana2011,C2,C4,C5}
\begin{eqnarray}
&&\hspace*{-1.4cm}\Phi(\xi,\tau)=\left[\psi_0~\mbox{sech}^2 \left(\frac{\xi-U\tau}{W}\right)\right]^\frac{1}{2}
\times \exp \left[\frac{i}{2P}\left\{U\xi+\left(\Omega_0-\frac{U^2}{2}\right)\tau \right\}\right],
\label{1eq:39}\\
&&\hspace*{-1.4cm}\Phi(\xi,\tau)=\left[\psi_0~\mbox{tanh}^2 \left(\frac{\xi-U\tau}{W}\right)\right]^\frac{1}{2}
\times \exp \left[\frac{i}{2P}\left\{U\xi-\left(\frac{U^2}{2}-2 P Q \psi_0\right)\tau \right\}\right],
\label{1eq:40}
\end{eqnarray}
where $\psi_0$ indicates the envelope amplitude, $U$ is the traveling speed
of the localized pulse, $W$ is the pulse width which can be written
as $W~[=(2P\psi_0/Q)^{1/2}]$, and $\Omega_0$ is the oscillating frequency
for $U = 0$. we have depicted bright and dark envelope solitons in Figs. \ref{1Fig:F4}(a)
and \ref{1Fig:F4}(b) by numerically analyzed Eqs. \eqref{1eq:39} and \eqref{1eq:40}, respectively.
\section{Conclusion}
\label{1sec:Conclusion}
The amplitude modulation of IAWs has been theoretically investigated in an unmagnetized four components PIP medium
consisting of positively and negatively charged inertial ions as well as inertialess $\kappa$-distributed electrons and positrons. A
NLSE, which governs the MI of IAWs and the formation of electrostatic bright and dark envelope solitons in PIP medium, is
derived by using the RPM. To conclude, we hope that our results may be useful in understanding the nonlinear phenomena (viz. MI of IAWs and envelope
solitons) in space PIP system (viz., upper regions of Titan's atmosphere \cite{El-Labany2012a},
chromosphere, solar wind, D-region ($\rm H^+, O_2^-$) and F-region ($\rm H^+, H^-$) of the
Earth's ionosphere \cite{Elwakil2010}, cometary comae \cite{Chaizy1991,Chowdhury2017})
and laboratory PIP system (viz., neutral beam sources \cite{Bacal1979}, plasma processing
reactors \cite{Gottscho1986}, and Fullerene ($\rm C^+, C^-$) \cite{Sabry2008,C8}, etc.).
\section*{Acknowledgements}
N. K. Tamanna is thankful to the Bangladesh Ministry of Science and Technology
for awarding the National Science and Technology (NST) Fellowship. A. Mannan
thanks the Alexander von Humboldt Foundation for a Postdoctoral Fellowship.


\begin{thebibliography}{99}
\bibitem{El-Labany2012a}S. K. El-Labany, W. M. Moslem \textit{et al.}, Astrophys. Space Sci. \textbf{338}, 3 (2012).

\bibitem{Elwakil2010}S. A. Elwakil, E. K. El-Shewy, and H. G. Abdelwahed, Phys. Plasmas \textbf{17}, 052301 (2010).

\bibitem{Chaizy1991}P. H. Chaizy, H. R\`{e}me, J. A. Sauvaud \textit{et al.}, Nature \textbf{349}, 393 (1991).

\bibitem{Chowdhury2017}N. A. Chowdhury, A. Mannan, M. M. Hasan, and A. A. Mamun, Chaos \textbf{27}, 093105 (2017).

\bibitem{Bacal1979}M. Bacal and G. W. Hamilton, Phys. Rev. Lett. \textbf{42}, 1538 (1979).

\bibitem{Gottscho1986}R. A. Gottscho and C. E. Gaebe, IEEE Trans. Plasma Sci. \textbf{14}, 92 (1986).

\bibitem{Sabry2008}R. Sabry,  Phys. Plasmas \textbf{15}, 092101 (2008).

\bibitem{Vasyliunas1968}V. M. Vasyliunas, J. Geophys. Res. \textbf{73}, 2839 (1968).

\bibitem{Sultana2011}S. Sultana and I. Kourakis,  Plasma Phys. Control. Fusion \textbf{53}, 045003 (2011).

\bibitem{Ghosh2012}D. K. Ghosh, P. Chatterjee, and B. Sahu,  Astrophys. Space Sci. \textbf{341}, 559 (2012).

\bibitem{Chatterjee2011}P. Chatterjee and U. N. Ghosh, Eur. Phys. J. D \textbf{64}, 413 (2011).

\bibitem{Saha2014}A. Saha, N. Pal, and P. Chatterjee,  Phys. Plasmas \textbf{21}, 102101 (2014).

\bibitem{El-Labany2012b}S. K. El-Labany, R. Sabry, W. F. El-Taibany, and E. A. Elghmaz,  Astrophys. Space Sci. \textbf{340}, 77 (2012).

\bibitem{C1} N. A. Chowdhury, A. Mannan, and A. A. Mamun, Phys. plasmas \textbf{24}, 113701 (2017).

\bibitem{C3} M. H. Rahman, A. Mannan, N. A. Chowdhury, and A. A. Mamun, Phys. Plasmas \textbf{25}, 102118 (2018).

\bibitem{C6} S. Jahan, N. A. Chowdhury, A. Mannan, and A. A. Mamun, Commun. Theor. Phys. \textbf{71}, 327 (2019).

\bibitem{Gharaee2011}H. Gharaee, S. Afghah, and H. Abbasi,  Phys. Plasmas \textbf{18}, 032116 (2011).

\bibitem{Chowdhury2018}N. A. Chowdhury, M. M. Hasan, A. Mannan, A. A. Mamun, Vacuum  \textbf{147}, 31 (2018).

\bibitem{Alinejad2014}H. Alinejad, M. Mahdavi, and M. Shahmansouri, Astrophys. Space Sci. \textbf{352}, 571 (2014).

\bibitem{Eslami2011}P. Eslami, M. Mottaghizadeh, and H. R. Pakzad, Phys. Plasmas \textbf{18}, 102313 (2011).

\bibitem{Ahmed2018}N. Ahmed, A. Mannan, N. A. Chowdhury, and A. A. Mamun, Chaos \textbf{28}, 123107 (2018).

\bibitem{Kourakis2003}I. Kourakis and P. K. Shukla, Phys. Plasmas \textbf{10}, 3459 (2003).

\bibitem{Kourakis2005} I. Kourakis and  P. K. Shukla, Nonlinear Proc. Geophys. \textbf{12}, 407 (2005).

\bibitem{C2} M. H. Rahman, N. A. Chowdhury, A. Mannan, M. Rahman, and A. A. Mamun, Chinese J. Phys. \textbf{56}, 2061 (2018).

\bibitem{C4} N. A. Chowdhury, A. Mannan, M. R. Hossen, and A. A. Mamun, Contrib. Plasma Phys. \textbf{58}, 870 (2018).

\bibitem{C5} N. A. Chowdhury, A. Mannan, M. M. Hasan, and A. A. Mamun, Plasma Phys. Rep. \textbf{45}, 01 (2019).

\bibitem{Dubinov2006}A. E. Dubinov, I. D. Dubinova, and V. A. Gordienko,   Phys. Plasmas \textbf{13}, 082111 (2006).

\bibitem{Dubinov2011}A. E. Dubinov, S. K. Saikov, and A. P. Tsatskin,  J. Exp. Theo. Phys. \textbf{112}, 1051 (2011).

\bibitem{Jannat2016}N. Jannat, M. Ferdousi, and A. A. Mamun, Plasma Phys. Rep. \textbf{42}, 678 (2016).

\bibitem{Saberian2017}E. Saberian, A. Esfandyri-Kalejahi, and M. Afsari-Ghazi, Plasma Phys. Rep. \textbf{43}, 83 (2017).

\bibitem{Kourakis2006}I. Kourakis, A. Esfandyri-Kalejahi, M. Mehdipoor, and P. K. Shukla, Phys. Plasmas \textbf{13}, 052117 (2006).

\bibitem{Esfandyri-Kalejahi2006a}A. Esfandyri-Kalejahi, I. Kourakis, M. Mehdipoor, and P. K. Shukla, J. Phys. \textbf{39}, 13817 (2006).

\bibitem{Esfandyri-Kalejahi2006b}A. Esfandyri-Kalejahi, I. Kourakis, and P. K. Shukla,  Phys. Plasmas \textbf{13}, 122310 (2006).

\bibitem{Abdelsalam2011}U. M. Abdelsalam, W. M. Moslem, A. H. Khater, and P. K. Shukla, Phys. Plasmas \textbf{18}, 092305 (2011).

\bibitem{El-Labany2011}S. K. El-Labany, E. K. El-Shewy \textit{et al.}, Plasma Phys. Rep. \textbf{18}, 092305 (2011).

\bibitem{C8}S.Khondaker \textit{et al.},  Contrib. Plasma Phys. e201800125. https://doi.org/10.1002/ctpp.201800125 (2019).

\end{thebibliography}
\end{document}